\documentclass[preprint,preprintnumbers,amsmath,amssymb]{revtex4}
\usepackage{graphicx}
\usepackage{dcolumn} 
\usepackage{bm} 
\begin{document}
\title{Quantum gravitational optics: the induced phase}
\author{ N.~Ahmadi$^{a}$\footnote{
Electronic address:~nahmadi@ut.ac.ir} and M.~Nouri-Zonoz $^{a,b}$ \footnote{
Electronic address:~nouri@theory.ipm.ac.ir}}
\address{$^{a}$ Department of Physics, University of Tehran, North Karegar Ave., Tehran 14395-547, Iran. \\
$^{b}$ Institute for studies in theoretical physics and mathematics, P O Box 19395-5531 Tehran, Iran.}
\begin{abstract}
The geometrical approximation of the extended Maxwell equation in curved spacetime incorporating interactions induced by the vacuum polarization effects is considered. Taking into account these QED interactions and employing the analogy between eikonal equation in geometrical optics and Hamilton-Jacobi equation for the particle motion, we study the phase structure of the modified theory. There is a complicated, local induced phase which is believed to be responsible for the modification of the classical picture of light ray. The main features of QGO could be obtained through the study of this induced phase. We discuss initial principles in conventional and modified geometrical optics and compare the results.
\end{abstract}
\maketitle
\section{Introduction}
Geometrical Optics (GO) is proved to be a useful approximation for the semiclassical study of photonic equation of motion, within the small wavelength limit. In fact the similarity between geometrical optics and particle dynamics has been the guiding principle to develop the semiclassical limit of particles motion, determined by the Hamilton-Jacobi equation. The phase (or eikonal) in GO plays the same role as the action of the particle in mechanics \cite{1}. Application of GO, to the electromagnetic wave propagation in a curved spacetime dates back to the studies initiated by Sachs (1961). Considering the photon as a quantum field and incorporating the coupling of the background curvature to the corresponding QED vacuum polarization effects modifies the corresponding action. Examining the reflection of this change on the GO formalism, leads to the introduction of a correction term (in eikonal equation and) to the photon wave vector. Vacuum polarization is an essential ingredient of QED whose contribution leads to astonishingly precise agreement between predicted and observed values of the electron magnetic moment and Lamb shift. Looking for its implications in the context of the so-called Quantum Gravitational Optics (QGO) is proved to be quite interesting and fruitful. It is found that a wide range of phenomena such as the polarization dependent propagation of photons as well as superluminal photon velocities become possible. These phenomena have been investigated in different gravitational backgrounds \cite{2}-\cite{3}-\cite{4}-\cite{5}-\cite{nut}.\\
Analyzing these phenomena through the mentioned analogy is the main interest in the present article. We interpret the above mentioned vacuum polarization effects through the modifications introduced by an induced phase. Generally, this phase is a complicated function of the local parameters appearing in the theory and therefore one could only identify its general properties. From the theoretical point of view, identifying the initial principles of GO i.e, the way the complex amplitude and the phase are separated, affects the results. We see that this even makes the mentioned effects to be revealed or neglected.\\
In the next section, the equation of motion appearing in QGO is discussed in the context of conventional GO. Applying the modified GO, as presented in \cite{6}, to the QGO equation of motion, we will compare the results in section III.
\section{Geometrical Optics approximation and Quantum gravitational optics}
As light propagates in a fixed curved background spacetime, its characteristics are subject to the laws of geometrical optics derived from the Maxwell equations in the presence of gravity. The GO solution is constructed through a perturbation method and results in the following laws \cite{gravitation}:
\begin{itemize}
\item light rays are null geodesics
\item The polarization vector is perpendicular to the ray and is parallelly propagated along the ray.
\end{itemize}
Mathematically, one exploits a GO solution with the assumption of a locally plane and monochromatic wave with a length scale that is very small compared to the typical curvature scale. The electromagnetic vector potential separates into a rapidly changing real phase $\theta$ and a slowly changing complex amplitude in the form of	
\begin{equation}
{\cal{A}}^{\mu}={\it{Re}}\left\{A^{\mu}+\epsilon B^{\mu}+...\right\}e^{i\theta/\epsilon}	
,\label{1}\end{equation}
where the parameter $\epsilon$ is introduced to keep track of the relative order of magnitudes of the terms included. In each small region of space, we can speak of a direction of propagation normal to a surface at all of whose points the phase of the wave is constant. We identify the wave vector as $k_{\mu}=\partial_{\mu}\theta$ and the polarization vector is introduced through $A_{\mu}=Aa_{\mu}$. The GO solution of the Maxwell's equations implies that the wave vector $k_{\mu}$ and the polarization vector $a_{\mu}$ specified at one point are fixed along the entire ray by their propagation equation. Since both propagation equations are nothing but parallel transport laws, the conditions $k^2=0$, $a^2=-1$ and $k.a=0$, once imposed on the vectors at one point will be satisfied along the entire ray.\\
The effect of one loop vacuum polarization on the Maxwell equations in a fixed curved background spacetime is represented by the following effective equation of motion derived by Drummond and Hathrell \cite{1}
\begin{equation}
D_{\mu}F^{\mu\nu}+\frac{1}{m^2}\left(2bR^{\mu \ \ }_{\ \ \lambda}D_{\mu}F^{\lambda\nu}+
4cR^{\mu\nu \ \ }_{ \ \ \lambda\rho}D_{\mu}F^{\lambda\rho}\right)=0
.\label{2}\end{equation}
Here, $b=\frac{13}{360}\frac{\alpha}{\pi}$, $c=
-\frac{1}{360}\frac{\alpha}{\pi}$ and in which $\alpha$ is the fine structure constant and $m$ is the electron 
mass. There are some approximations under which this equation of motion was obtained. The first one is the low 
frequency approximation in the sense that the derivation is only applicable to wavelengths 
$\lambda>\lambda_{c}$. By this approximation we ignore terms in the effective action involving 
higher order field derivatives. The second is a weak field approximation for gravity which 
implicitly means that the typical curvature scale, $L$ is much larger than the electron Compton wavelength, i.e, $\lambda_c \ll L$.\\
Considering the one loop vacuum polarization affecting the photon propagation in a fixed local background, the above view shifts slightly. The null cone and phase velocity modifications are resulted as a direct consequence of assigning the GO solution to the modified Maxwell equations in a curved spacetime. The two length-scale expansion is typical of quantum field theory calculation in a background field. The general form of the perturbative expansion in the coupling constant is $ f_{0}\left(\epsilon\right)+\alpha f_{1}\left(\epsilon\right)+\alpha^{2}f_{2}\left(\epsilon\right)+...$, where $f_{i}\left(\epsilon\right)$ are functions of a dimensionless parameter $\epsilon$ defined as the ratio of the physical length scale $\lambda_{c}$ to the background scale $L$, i.e, $\frac{\lambda_{c}}{L}$. For the present calculation, an expansion of $f_{i}\left(\epsilon\right)$ for small $\epsilon$, like what is given in (\ref{1}), is inserted in eq. (\ref{2}), in other words only terms up to $O(\alpha)$ are taken into account.
 In this notation, $D_{\lambda}F^{\mu\nu}$ can be written as 
\begin{eqnarray}
D_{\lambda}F^{\mu\nu}&=&{\it{Re}}\left\{\left[-\frac{1}{\epsilon^2}\left[k^{\nu}k_{\lambda}\left(A^{\mu}+\epsilon B^{\mu}+...\right)-k^{\mu}k_{\lambda}\left(A^{\nu}+\epsilon B^{\nu}+...\right)\right]\right.\right.\nonumber\\
&&+\frac{i}{\epsilon}k_{\lambda}\left[\left(A^{\mu}+\epsilon B^{\mu}+...\right)^{;\nu}-\left(A^{\nu}+\epsilon B^{\nu}+...\right)^{;\mu}\right]+\frac{i}{\epsilon}\left[k^{\nu}_{\ ;\lambda}\left(A^{\mu}+\epsilon B^{\mu}+...\right)\right.\nonumber\\
&&\left.-k^{\mu}_{\ \ ;\lambda}\left(A^{\nu}+\epsilon B^{\nu}+...\right)\right]+\frac{i}{\epsilon}\left[k^{\nu}\left(A^{\mu}+\epsilon B^{\mu}+...\right)_{;\lambda}-k^{\mu}\left(A^{\nu}+\epsilon B^{\nu}+...\right)_{;\lambda}\right]\nonumber\\
&&\left.\left.+\left(A^{\mu}+\epsilon B^{\mu}+...\right)^{;\nu}_{\ \ ;\lambda}-\left(A^{\nu}+\epsilon B^{\nu}+...\right)^{;\mu}_{\ \ ;\lambda}\right]e^{i\frac{\theta}{\epsilon}}\right\}.
\label{expansion}\end{eqnarray}
After applying the expansion eq. (\ref{expansion}) to the equation of motion, the next step is to collect terms of order $\frac{1}{\epsilon^2}$ and $\frac{1}{\epsilon}$ (terms of order higher than $\frac{1}{\epsilon}$ govern post-geometric corrections). The leading term, $O\left(\frac{1}{\epsilon^2}\right)$ result in the modified light cone as 
\begin{equation}
k^{2}A^{\nu}-k^{\nu}k_{\mu}A^{\mu}+\frac{1}{m^2}\left[2bR^{\mu}_{\ \lambda}\left(-k^{\nu}k_{\mu}A^{\lambda}+k_{\mu}k^{\lambda}A^{\nu}\right)+4cR^{\mu\nu}_{\ \ \lambda\rho}\left(-k_{\mu}k^{\rho}A^{\lambda}+k^{\lambda}k_{\mu}A^{\rho}\right)\right]=0
\label{3}\end{equation}
which for transverse photons with polarization vector normalized to unity, reduces to the following equation 
\begin{equation}
k^2+\frac{2b}{m^2}R_{\mu\lambda}k^{\mu}k^{\lambda}-\frac{8c}{m^2}R_{\mu\nu\lambda\rho}
k^{\mu}k^{\lambda}a^{\nu}a^{\rho}=0.
\label{4}\end{equation} 
The subleading term, $O(\frac{1}{\epsilon})$, in the equation of motion gives
\begin{eqnarray}
&&-i\left\{\left(-k^{\nu}k_{\mu}B^{\mu}+k^{\mu}k_{\mu}B^{\nu}\right)+\frac{1}{m^2}\left[2bR^{\mu}_{\ \lambda}\left(-k^{\nu}k_{\mu}B^{\lambda}+k^{\lambda}k_{\mu}B^{\nu}\right)+4cR^{\mu\nu}_{\ \ \lambda\rho}\left(-k^{\rho}k_{\mu}B^{\lambda}+k^{\lambda}k_{\mu}B^{\rho}\right)\right]\right\}\nonumber\\
&&+\left[k_{\mu}\left(A^{\mu;\nu}-A^{\nu;\mu}\right)+k^{\nu}_{\ ;\mu}A^{\mu}-k^{\mu}_{\ ;\mu}A^{\nu}+k^{\nu}A^{\mu}_{\ ;\mu}-k^{\mu}A^{\nu}_{\ ;\mu}\right]\nonumber\\
&&+\frac{1}{m^2}\left\{2bR^{\mu}_{\ \lambda}\left[k_{\mu}\left(A^{\lambda;\nu}-A^{\nu;\lambda}\right)+k^{\nu}_{\ ;\mu}A^{\lambda}-k^{\lambda}_{\ ;\mu}A^{\nu}+k^{\nu}A^{\lambda}_{\ ;\mu}-k^{\lambda}A^{\nu}_{\ ;\mu}\right]\right.\nonumber\\
&&\left.+4cR^{\mu\nu}_{\ \ \lambda\rho}\left[k_{\mu}\left(A^{\lambda;\rho}-A^{\rho;\lambda}\right)+\left(k^{\rho}_{\ ;\mu}A^{\lambda}-k^{\lambda}_{\ ;\mu}A^{\rho}\right)+\left(k^{\rho}A^{\lambda}_{\ ;\mu}-k^{\lambda}A^{\rho}_{\ ;\mu}\right)\right]\right\}=0
\end{eqnarray}
For travsverse photon and after employing equation (\ref{3}), it transforms into:
\begin{eqnarray}
\nabla_{k}A^{\nu}=\frac{1}{2}\left(k^{\nu}A^{\mu}_{\ ;\mu}-k^{\mu}_{\ ;\mu}A^{\nu}\right)+\frac{1}{m^2}\left\{bR_{\mu\lambda}\left[k^{\nu}A^{\lambda;\mu}-k^{\lambda}A^{\nu;\mu}+\left(k^{\mu}A^{\lambda}\right)^{;\nu}-\left(k^{\mu}A^{\nu}\right)^{;\lambda}\right.\right]\nonumber\\+\left.\left(4c\right)R_{\mu\nu\lambda\rho}\left[\left(k^{\rho}A^{\lambda}\right)^{;\mu}+k^{\mu}A^{\lambda;\rho}\right]\right\}	
,\label{7}\end{eqnarray}
which determines the first order variation of ${A}^{\mu}=Aa^{\mu}$ as the vector amplitude along the ray.\\
Eq. (\ref{4}) is an effective light cone equation, representing the wave vector changes induced by QED interactions. It shows that at this level of approximation, the wave vector acquires an additional polarization dependent component defined as
\begin{eqnarray}
k_{\mu}&=&k_{\mu}^{\left(0\right)}-\frac{1}{m^2}\left[bR_{\mu\lambda}k^{\lambda}
-\left(4c\right)R_{\mu \lambda\sigma\kappa}a^{\lambda}a^{\kappa}k^{\sigma}\right]\nonumber\\
&\equiv&k_{\mu}^{\left(0\right)}+\partial_{\mu}\Phi.
\label{5}\end{eqnarray}
 $\Phi$ can be understood as a single-valued {\it{local}} phase of order $\alpha$, which like any other first order correction term, must be calculated along the zeroth-order approximation (i.e, with $k=k^{\left(0\right)}$ and $a=a^{\left(0\right)}$). Following the definition of $k_{\mu}$ as a gradient satisfying $k_{\mu;\nu}=k_{\nu;\mu}$, one can obtain the photon trajectories corresponding to the new equation of motion (\ref{2}) as 
\begin{eqnarray}
\nabla_{k}k^{\nu}&=&k^{\mu}D^{\nu}k_{\mu}=\frac{1}{2}D^{\nu}k^{2}=\nonumber\\
&=&-\frac{1}{m^2}D^{\nu}\left[bR_{\mu\lambda}k^{\mu}k^{\lambda}
-\left(4c\right)R_{\mu \lambda\sigma\kappa}a^{\lambda}a^{\kappa}k^{\mu}k^{\sigma}\right]	
.\label{6}\end{eqnarray}
 Here the bracket times $-\frac{1}{m^2}$ in the second line can be identified as $\nabla_{k}\Phi$, which is not generally zero. 
 Note that in the framework of the conventional geometrical optics, the phase characteristics of the potential vector are now determined by the phase $\theta +\Phi$ and the equation (\ref{4}), whereas the vector amplitude $A^{\mu}=Aa^{\mu}$ and consequently the polarization vector (of the zeroth-order in $\epsilon$) are specified by the equation (\ref{7}). However as long as we work with equation (\ref{6}), the wave and the polarization vectors in right hand side are the zeroth-order (classical) quantities and we will consider manifestations of the induced phase for a photon in a state of adiabatically invariant polarization vector.\\ The phase introduced in (\ref{5}) also affects the propagation of a bundle of rays. For infinitesemally nearby geodesics in the family, the geodesic deviation equation relates the relative acceleration to the Riemann tensor. Let $\xi^{\mu}$ be the deviation vector normal to the wave vector which connects abreast rays, so that $\xi.k=0$ and $\nabla_{k}\xi=\nabla_{\xi}k$. The relative velocity of two infinitesimally nearby geodesics is given by $v^{\mu}=\nabla_{k}\xi^{\mu}$ and the relative acceleration is
\begin{equation}
{\textsl{a}}^{\mu}\equiv \nabla_{k}\nabla_{k}\xi^{\mu}=R_{\nu\lambda\sigma}^{\ \ \ \ \mu}k^{\nu}k^{\sigma}\xi^{\lambda}
.\label{8}\end{equation}
The role of the induced phase, $\Phi$, shows up in the study of the perturbative deformation of the bundle's cross section where effective counterparts for optical scalars namely effective expansion, shear and vorticity are introduced in the study of Raychaudhuri equations in QGO \cite{10}.
In summary the following general characteristics could be obtained from the above first order QGO corrections in the context of GO approximation,
\begin{itemize}
\item The polarization degeneracy has been removed.
\item Depending on the direction and polarization of photons, superluminal propagation becomes possible.
\item The presence of superluminal photons and interactions violating the strong equivalence principle does not necessarily imply causality violation, although there is no reason to prove that causality violation does not occur.
\item The photons with classical polarizations acquire equal phases but with different signs. This in turn gives rise to two opposite trajectories both departed from the zero-order one.
\item Due to the polarization sum rule, the sum of averaged velocity shifts for two physical polarizations is zero in Ricci flat spacetimes \cite{7}. 
\item Both the wave vector and the polarization vector are transported along the ray according to the equations (\ref{6}) and (\ref{7}), respectively. Since any polarization is possible in quantum theory, the final observed state may not retain the transverse nature of the wave.  
\item Besides the classical parameters, some quantum corrections of O($\alpha$), depending on the change in the curvature along a geodesic may affect the focusing, twist and shearing of a bundle of rays and equally the associated Raychaudhuri equation \cite{10}.
	\end{itemize} 
\section{QGO and the modified GO} 
We have outlined the quantum corrections to the propagation characteristics of a photon in a fixed gravitational background through the conventional GO approximation. The relative order of magnitude for these modifications is $O\left(\frac{\alpha\lambda_{c}^2}{L^2}\right)$, which is extremely small. We attributed corrections to the photon wave vector (or a phase) induced by the interactions. Multiplying the equation (\ref{6}) by $k_{\nu}$, we get
\begin{equation}
k_{\nu}\nabla_{k}k^{\nu}
=\nabla_{k}\left[\frac{-1}{m^2}\left(bR_{\mu\lambda}k^{\mu}k^{\lambda}
-\left(4c\right)R_{\mu \lambda\sigma\kappa}a^{ \lambda}a^{\kappa}k^{\mu}k^{\sigma}\right)\right]=O\left(\alpha^2\right)	
.\label{9}\end{equation}
This means that
\begin{equation}
k_{\nu}\nabla_{k}k^{\nu}=\nabla_{k}\nabla_{k}\Phi\approx0
.\label{10}\end{equation}
This calculation shows that the first order correction to the wave vector is orthogonal to its zeroth-order direction. The gradient of a polarization dependent phase is a function of the local values of parameters, so cannot exceed $O\left(\alpha\right)$. In this interpretation the effects (velocity shift and trajectory splitting) listed in the previous section are the results of this polarization dependant phase deviation and follow immediately from the initial principles of GO. We recall that in the conventional GO, the phase characteristics of the wave vector are determined by the phase $\theta$ and its evolution is governed by the null cone equation, whereas the $A^{\mu}$ specifies the amplitude and the polarization. In this framework, interactions could alter the phase up to the relevant order. This way of separation, however, is not unique. In a modified version of GO, suggested by Bliokh and Bliokh \cite{6}, the complex amplitude and the phase of the electromagnetic field are separated in the following manner,
\begin{equation}
{\cal{A}}^{\mu}=\widehat{A}^{\mu}e^{i\theta	/\epsilon},\ \ \ \ \ \  \widehat{A}^{\mu}=Aa^{\mu}e^{i\Phi/\epsilon}
.\label{11}\end{equation}
The phase is separated into the local phase $\Phi$ and the non-local phase $\theta$. The wave vector is defined through the gradient of the non-local part, $k_{\mu}=\partial_{\mu}\theta$, while the polarization is specified by the amplitude eigenvector. There is an ambiguity in this separation which is determined up to the gauge transformation
\begin{equation}
\theta\rightarrow\theta -\Phi^{'}\ \ \ \ \ \ \ \ \Phi\rightarrow\Phi+\Phi^{'}
.\label{12}\end{equation}
Here $\Phi^{'}$ is a local phase. It is believed that these transformations have no effect on the physically observable quantities. The phase increment is dertermined by the entire path covered by the wave, so it is conceptually a non-local or integral quantity, contrary to the amplitude which is a local quantity. \\
The wave vector deviation induced by the interactions, can be associated with the gradient of either of these phases, local or non-local. If the deviation has no projection along its zeroth-order component (transverse deviation), it is the gradient of a local phase. So it could be part of the amplitude, under a gauge transformation, with a slight distortion of the phase front. It generally has a small value which does not exceed the particle wavelength. In view of the uncertainty relation, the wave vector is not a physically measurable quantity in this range.
Interactions may induce longitudinal wave vector deviations if the induced phase is a non-local one. These deviations can not change the direction of the normal to the phase front and their integral after the transport along the ray does not vanish.\\
An example of the first group of deviations is the one we encounter in the study of QED interactions in a curved background formulated as QGO through the Drummond-Hathrell \cite{1} action \footnote{Note that we have $k_{\nu}\nabla_{k}\partial^{\nu}\Phi=0$ as it is mentioned in the paragraph below the equation (\ref{10}).} or those in an arbitrary anisotropic (but homogenous) electromagnetic field described by the Euler-Heisenberg action \cite{7}.
The second familiar group of phase deviations is that of the electron wave vector in the presence of a topologically non-trivial electromagnetic field (in the form of minimal substitution) leading to the spectacular quantum interference phenomena, known as the Bohm-Aharanov effect. At the heart of these group of interactions lies a non-integrable phase  $\oint_{c}A_{\mu}dx^{\mu}$ which arises after the wave function transport around a closed path $c$. In quantum mechanics there exists an analogous non-local topological phase, the so called Berry's phase, arising from the transport around a closed path in momentum space in the presence of a gauge field or in a smooth inhomogenous medium.\\
So according to this modified version of GO,
\begin{itemize}
\item the induced phase can be attributed to the change of amplitude and
\item the ray shifts would be related to the uncertainty in determining the ray trajectory within the scale of the theoy. 
\end{itemize}
So there is no observable physical effect in QGO as long as our observables depend on the ray trajectory and the phase (the integral of the wave vector on the ray). In other words, since the deviations are within the range of uncertainty principle the observablity seems to be an ambiguous concept. Also if the classical theory respects the causality, there is no reason to doubt that the QED in curved spacetime remains a causal theory. 
\section{discussion}
In the present article we have considered the characteristics of the QED theory in which the interactions of vacuum polarization effects with the curvature are taken into account. We rely on the analogy between the eikonal equation in geometric optics and the particle equation of motion. The analogy is exploited to find the phase structure and the trajectory of the propagating photon, semiclassically. (In this way, the photon is treated as a test particle, so its effect on the metric is assumed to be negligible). It turns out that interpreting the results, at least at the first order geometric optics, depends to a large extent on the way the amplitude and the phase of the electromagnetic wave are separated. The concepts of local (non-integrable) and non-local (integrable) phases are already proved to be important and the QED effects (the shift in phase velocity and ray trajectory) could be interpreted differently, in conventional and modified geometric optics. There are some guiding points which convinces one to rely on the modified GO expectancies: First, in \cite{5} it is shown that, the cosmological constant, $\Lambda$, and the the topological structure parameter, $k$, are failed to enter the velocity shift, secondly, in \cite{10} we have discussed the invariant quantities in the QGO which reduce to their classical counterparts in the limit of zero perturbation and therefore there remains no anomaly in the theory. Effective Raychaudhuri equation, also discussed in \cite{10}, has the general form of its classical version. Finally, as we have seen in (\ref{8}) the geodesic deviation vector in modified theory satisfies the same (Jacobi) equation as the classical deviation vector does. It seems that whenever we touch upon the evolution equations for the {\it{effective}} quantities, there is no difference between them and their classical counterpart. This is not surprising, since $\nabla_{k}\nabla_{k}\Phi=0$ and there are no basic changes, at least to this order of approximation. These facts support the non-integrable nature of QGO phase. In literature \cite{9}, the matter of observability is linked to the limiting condition $\frac{\alpha \lambda_{c}^{2}}{\lambda L}>1$ i.e, the length difference between paths corresponding to different polarizations, $\frac{\alpha \lambda_{c}^{2}}{L}$, should be larger than the photon wavelength, $\lambda$. This length discrepancy is given by $\delta s\approx L\delta v\approx \frac{\alpha \lambda_{c}^{2}}{L}$, where the essential assumption is that the first order correction to the velocity must be along its zeroth order direction, contrary to the results mentioned in this paper. However we believe that due to the uncertainty relation and the non-integrablity of QGO phase, these interactions can not be detected.
\section *{Acknowledgments} 
The authors would like to thank University of Tehran for supporting this project under the grants provided by the research council.


\begin{references}
\bibitem{1} L. D. Landau and E. M. Lifshitz, The classical theory of fields, Pergamon press, 1975.
\bibitem{2} I. T. Drummond and S. J. Hathrell, Phs. Rev. D22 (1980) 343.
\bibitem{3} R. D. Daniels and G. M. Shore, Nucl. Phys. B425 (1994) 634.
\bibitem{4} R. D. Daniels and G. M. Shore, Phys. Lett. B367 (1996) 75-83.
\bibitem{5} Rong-Gen Cai, Nucl. Phys. B524, (1998) 639.
\bibitem{nut} N. Ahmadi, S. Khoeini-Moghaddam and M. Nouri-Zonoz, arxiv:gr-qc/0607135.
\bibitem{gravitation} C. W. Misner, K. S. Thorne and J. A. Wheeler, Gravitation, Freeman, San Francisco, 1973.
\bibitem{6} K. Yu. Bliokh and Yu. P. Bliokh, Phys. Rev. E70, (2004) 026605
\bibitem{7} G. M. Shore, Nucl. Phys. B460 (1996) 379.
\bibitem{8} G. M. Shore, Nucl. Phys. B605 (2001) 455.
\bibitem{9} G. M. Shore, Contemp. Phys. 44 (2003) 503.
\bibitem{10} N. Ahmadi and M. Nouri-Zonoz, Phys. Rev. D74 (2006)44034.
\end{references}
\end{document}